\documentclass{ws-procs11x85}
\usepackage{balance} 

\newcommand{\met}{\hbox{E\kern-0.5em\lower-0.1ex\hbox{/}}_T}

\begin{document}

\twocolumn[
\title{Notes on Translational and  Rotational  Properties
of Tensor Fields in Relativistic Quantum Mechanics}\thanks{Presented also at the FFP14, Marseille, Jul. 15-18, 2014 and the ICCA-10, Tartu, Estonia, Aug. 4-9, 2014.}

\author{V. V. Dvoeglazov}

\address{UAF, Universidad de Zacatecas\\
Apartado Postal 636, Suc. 3 Cruces, Zacatecas 98068 Zac., M\'exico\\
Web: http://fisica.uaz.edu.mx/\~{}valeri/\\
E-mail: valeri@fisica.uaz.edu.mx}


\begin{abstract}

Recently, several discussions on the possible observability
of 4-vector fields have been published in literature.
Furthermore, several authors recently claimed existence of
the helicity=0 fundamental field.
We re-examine the theory of antisymmetric tensor
fields and 4-vector potentials. We study the massless limits.
In fact, a theoretical motivation for this venture is the old papers of
Ogievetski\u{\i} and Polubarinov, Hayashi, and Kalb and Ramond.
Ogievetski\u{\i} and Polubarinov proposed
the concept of the {\it notoph}, whose helicity
properties are complementary to those of the {\it photon}.
We analyze the quantum field theory with taking into account 
mass dimensions of the notoph and the photon. 
It appears to be  possible to describe both photon and
notoph degrees of freedom on the basis of
the modified Bargmann-Wigner formalism for the symmetric
second-rank spinor. 
Next, we proceed to derive equations for the symmetric tensor
of the second rank on the basis of the Bargmann-Wigner formalism
in a straightforward way. The symmetric multispinor of the fourth
rank is used. Due to serious problems with the interpretation of
the results obtained on using the standard procedure we generalize
it and obtain the spin-2 relativistic equations, which are consistent
with the general relativity. 
Thus, in fact we deduced the gravitational field equations from 
relativistic quantum mechanics. 
The relations of this theory with the scalar-tensor theories 
of gravitation and f(R) are discussed. Particular attention has been paid to 
the correct definitions of the energy-momentum tensor and other N\"other currents
in the electromagnetic theory,
the relativistic theory of gravitation, the general relativity, and their generalizations.
We estimate possible interactions, fermion-notoph,
graviton-notoph, photon-notoph, and  we conclude that they can probably be seen 
in experiments  in the next few years.\\
PACS number: 03.65.Pm , 04.50.-h , 11.30.Cp
\end{abstract}
\keywords{Notoph; Graviton}
\vskip12pt  
]

\bodymatter

\section{Introduction}

In this presentation we re-examine the theory of   the 4-potential field, the antisymmetric
tensor fields of the second ranks and the spin-2 fields coming from the modified Bargmann-Wigner formalism. 
In the series of the papers~\cite{Ahlu,DVALF,DVO962,DVO-HJS}
we tried to find connection between the theory
of the quantized antisymmetric tensor (AST) field of the second rank (and that
of the corresponding 4-vector field) with the $2(2s+1)$ Weinberg-Tucker-Hammer
formalism~\cite{Weinberg,TH1}. 
Several previously published works, Refs.~\cite{Ogievet,Hayashi,Kalb,DEB,Avdeev}, 
introduced the concept of the {\it notoph} 
(the Kalb-Ramond field) which is constructed on the basis of 
the antisymmetric tensor ``potentials". It represents itself the non-trivial 
spin-0 field. The well-known textbooks~\cite{Bogol,Novozh,Itzyk} did {\it not}
discuss  the problems, whether the massless {\it quantized} AST ``potential"
and the {\it quantized} 4-vector field are transverse or longitudinal fields
(in the sense if the helicity $h=\pm 1$ or $h=0$)? can the electromagnetic potential 
be a 4-vector in a quantized theory ?  contradictions with the Weinberg theorem ``that no symmetric
tensor field of rank $s$ can be constructed from the creation and
annihilation operators of massless particles of spin $s$"? how should 
the massless limit be taken? and many other fundamental problems of the physics of bosons. 
However, one can advise Refs.~\cite{Weinberg,Wein}.

First of all, after a referee of ``Found. Phys." and ``Int. J. Mod. Phys." 
we note that 1) ``...In natural units
($c=\hbar=1$) ... a lagrangian density, since the action is
dimensionless,
has dimension of [energy]$^4$"; 2) One can always renormalize the
lagrangian density and ``one can obtain the same equations of motion...
by substituting $L \rightarrow (1/M^N) L$, where $M$ is an arbitrary
energy scale",  cf.~Ref.~\cite{DVO962}; 3) the right physical dimension of
the field strength tensor $F^{\mu\nu}$ is [energy]$^2$; ``the transformation
$F^{\mu\nu} \rightarrow (1/2m) F^{\mu\nu}$ [which was regarded in
Ref.~\cite{DVO97,physscripta}] ... requires a more detailed study ... [because] the
transformation above changes its physical dimension: it is not a simple
normalization transformation". Furthermore, in the first papers on the
notoph~\cite{Ogievet,Hayashi,Kalb} the authors
used the normalization of the 4-vector $F^\mu$ field (which  is related to a third-rank antisymmetric field tensor) 
to [energy]$^2$ and, hence, the antisymmetric tensor ``potentials"
$A^{\mu\nu}$, to [energy]$^1$. We try to discuss these problems on the basis of the
generalized Bargmann-Wigner formalism~\cite{BW}. The Proca and Maxwell formalisms
are generalized too, see, e.~g., Ref.~\cite{DVO-HJS}.
In the next Sections we consider the spin-2 equations. 
A field of the rest mass $m$ and the spin $s \geq
{1\over 2}$ is represented by a completely symmetric multispinor of rank $2s$.
The particular cases $s=1$ and $s={3\over 2}$ have been considered in the
textbooks, e.~g., Ref.~\cite{Lurie}. The spin-2 case can also be of some
interest. Nevertheless,
questions of the redundant components of the higher-spin relativistic
equations are not yet understood in detail~\cite{Kirch}.
In the last Sections (in the previous papers of us, as well) we discuss the questions of quantization, 
interactions and relations between various higher-spin theories.

\section{4-potentials and Antisymmetric Tensor Field. Normalization.}

The spin-0 and spin-1 field particles can be constructed by taking the direct
product of 4-spinors~\cite{BW,Lurie}.
Let us firstly repeat the Bargmann-Wigner procedure  for bosons of spin 0 and 1.
The set of basic equations for  $s=0$ and $s=1$ are written, e.g., in
Ref.~\cite{Lurie}
\begin{eqnarray} \left [
i\gamma^\mu \partial_\mu -m \right ]_{\alpha\beta} \Psi_{\beta\gamma} (x)
&=& 0\,,\label{bw1}\\ \left [ i\gamma^\mu \partial_\mu -m \right
]_{\gamma\beta} \Psi_{\alpha\beta} (x) &=& 0\,.\label{bw2}
\end{eqnarray}
We expand the $4\times 4$ matrix field function into the antisymmetric and
symmetric parts in the standard way
\begin{eqnarray}
\Psi_{[\alpha\beta ]} &=& R_{\alpha\beta} \phi +
\gamma^5_{\alpha\delta} R_{\delta\beta} \widetilde \phi +
\gamma^5_{\alpha\delta} \gamma^\mu_{\delta\tau} R_{\tau\beta} \widetilde
A_\mu \,,\label{as}\\ \Psi_{\left \{ \alpha\beta \right \}} &=&
\gamma^\mu_{\alpha\delta} R_{\delta\beta} A_\mu
+\sigma^{\mu\nu}_{\alpha\delta} R_{\delta\beta} F_{\mu\nu}\,,
\label{sisi}
\end{eqnarray}
where $R= CP$ has the
properties (which are necessary to make expansions (\ref{as},\ref{sisi}) to
be possible in such a form)
\begin{eqnarray}
&& R^T = -R\,,\, R^\dagger =R = R^{-1}\,,\\
&& R^{-1} \gamma^5 R = (\gamma^5)^T\,,\\
&& R^{-1} \gamma^\mu R = -(\gamma^\mu)^T\,,\\
&& R^{-1} \sigma^{\mu\nu} R = - (\sigma^{\mu\nu})^T\,.
\end{eqnarray}
The  explicit form of this matrix can be chosen:
\begin{equation}
R=\begin{pmatrix}i\Theta & 0\\ 0&-i\Theta\end{pmatrix}\,,\, \Theta = -i\sigma_2 =
\begin{pmatrix}0&-1\\ 1&0\end{pmatrix},
\end{equation} provided that $\gamma^\mu$
matrices are in the Weyl representation.  The equations
(\ref{bw1},\ref{bw2}) lead to the Kemmer set of the $s=0$ equations:
\begin{eqnarray}
m\phi &=&0 \,,\\
m\widetilde \phi &=& -i\partial_\mu \widetilde A^\mu\,,\\
m\widetilde A^\mu &=& -i\partial^\mu \widetilde \phi\,,
\end{eqnarray}
and to the Proca-Duffin-Kemmer set of the equations for the $s=1$
case:\,\,\footnote{We could use another symmetric matrix $\gamma^5
\sigma^{\mu\nu} R$ in the expansion of the symmetric spinor of the second
rank~\cite{physscripta}.  In this case the equations are
\begin{eqnarray}
&& i\partial_\alpha \widetilde F^{\alpha\mu} +{m\over 2} B^\mu = 0\,,
\label{de1}\\
&& 2im \widetilde F^{\mu\nu} = \partial^\mu B^\nu -\partial^\nu
B^\mu\,,\label{de2}
\end{eqnarray}
in which  the dual tensor
$\widetilde F^{\mu\nu}= {1\over 2} \epsilon^{\mu\nu\rho\sigma}
F_{\rho\sigma}$ presents,
because we used that in the Weyl representation
$\gamma^5 \sigma^{\mu\nu} = {i\over 2} \epsilon^{\mu\nu\rho\sigma}
\sigma_{\rho\sigma}$; $B^\mu$ is the corresponding vector potential.  The
equation for the antisymmetric tensor field (which can be obtained from
this set) does not change its form  but we see some
``renormalization" of the field functions. In general, it is permitted to
choose various relative  factors in the expansions of the
symmetric wave function (4). It is also permitted to consider the matrix term of the form
$\gamma^5 \sigma^{\mu\nu}$.  We have additional factors in
equations connecting  physical fields and their potentials.  They
can be absorbed by  redefinitions of the potentials/fields.
The above shows that the dual tensor of the
second rank can also be epxanded in potentials. See  below.}
\begin{eqnarray} &&\partial_\alpha
F^{\alpha\mu} + {m\over 2} A^\mu = 0 \,,\label{1} \\ &&2 m F^{\mu\nu} =
\partial^\mu A^\nu - \partial^\nu A^\mu \,, \label{2} \end{eqnarray} In
the meantime, in the textbooks, the latter set is usually written as ({\it
e.g.}, p. 135 of Ref.~\cite{Itzyk})
\begin{eqnarray}
&&\partial_\alpha F^{\alpha\mu} + m^2 A^\mu = 0 \,, \label{3}\\ &&
F^{\mu\nu} = \partial^\mu A^\nu - \partial^\nu A^\mu \,, \label{4}
\end{eqnarray} The set (\ref{3},\ref{4}) is obtained from
(\ref{1},\ref{2}) after the normalization change $A_\mu \rightarrow 2m
A_\mu$, or $F_{\mu\nu} \rightarrow {1\over 2m} F_{\mu\nu}$.  Of course, one
can investigate other sets of equations with different normalizations of
the $F_{\mu\nu}$ and $A_\mu$ fields. Are all these sets of equations
equivalent?  As we can see, to answer this question is not trivial.
At the moment, we mention that the massless limit can and must be taken
in the end of  calculations only, {\it i.~e.}, for physical quantities.

In order to be able to answer the question about
the behaviour of eigenvalues of the spin operator
${\bf J}^i = {1\over 2} \epsilon^{ijk}
J^{jk}$ in the massless limit
one should know the behaviour of the fields $F_{\mu\nu}$ and/or $A_\mu$ in
the massless limit.  We want to analyze the first set (\ref{1},\ref{2}).
If one chooses the following definitions (p. 209 of Ref.~\cite{Wein})
\begin{eqnarray}
&&\epsilon^\mu  ({\bf 0}, +1) = - {1\over \sqrt{2}}
\begin{pmatrix}0\cr 1\cr i \cr 0\end{pmatrix}\,,\,
\epsilon^\mu  ({\bf 0}, 0) =
\begin{pmatrix}0\cr 0\cr 0 \cr 1\end{pmatrix}\,,\,\\
&&\epsilon^\mu  ({\bf 0}, -1) = {1\over \sqrt{2}}
\begin{pmatrix}0\cr 1\cr -i \cr 0\end{pmatrix}\,,
\end{eqnarray}
and ($\widehat p_i = p_i /\mid {\bf p} \mid$,\, $\gamma
= E_p/m$), p. 68 of Ref.~\cite{Wein},
\begin{eqnarray} && \epsilon^\mu ({\bf p}, \sigma) =
L^{\mu}_{\,\nu} ({\bf p}) \epsilon^\nu ({\bf 0},\sigma)\,, \\
&& L^0_{\, 0} ({\bf p}) = \gamma\, ,\, L^i_{\, 0} ({\bf p}) =
L^0_{\, i} ({\bf p}) = \widehat p_i \sqrt{\gamma^2 -1}\, ,\\
&&L^i_{\, k} ({\bf p}) = \delta_{ik} +(\gamma -1) \widehat p_i \widehat
p_k \, \end{eqnarray}
for the 4-vector potential field,\footnote{Remember that the invariant
integral measure over the Minkowski space for physical particles is $$\int
d^4 p \delta (p^2 -m^2)\sim \int {d^3  {\bf p} \over 2E_p}\,,\,
E_p = \sqrt{{\bf p}^2 +m^2}\,.$$ Therefore, we use the field operator
as in (\ref{fo}). The coefficient $(2\pi)^3$ can be considered at this
stage as chosen for the convenience.  In Ref.~\cite{Wein} the factor
$1/(2E_p)$ was absorbed in creation/annihilation operators, and instead of
the field operator (\ref{fo}) the operator was used in which the
$\epsilon^\mu ({\bf p}, \sigma)$ functions for a massive spin-1 particle
were substituted by $u^\mu ({\bf p}, \sigma) = (2E_p)^{-1/2} \epsilon^\mu
({\bf p}, \sigma)$.}
p. 129 of Ref.~\cite{Itzyk}
\begin{eqnarray} 
A^\mu (x^\mu) =
&&\sum_{\sigma=0,\pm 1} \int {d^3 {\bf p} \over (2\pi)^3 }
{1\over 2E_p} \left
[\epsilon^\mu ({\bf p}, \sigma) a ({\bf p}, \sigma) e^{-ip\cdot x} +\right .\nonumber\\
&+&\left . (\epsilon^\mu ({\bf p}, \sigma))^c b^\dagger ({\bf p}, \sigma) e^{+ip\cdot
x} \right ]\,, \label{fo}
\end{eqnarray}
The normalization of the wave
functions in the momentum representation is chosen to the unit,
$\epsilon_\mu^\ast ({\bf p}, \sigma) \epsilon^\mu ({\bf p},\sigma) = -
1$.\footnote{The metric used in this paper $g^{\mu\nu} = \mbox{diag} (1,
-1, -1, -1)$ is different from that of Ref.~\cite{Wein}.}$^,$\footnote{In this paper 
we assume that $[\epsilon^\mu ({\bf
p},\sigma) ]^c =e^{i\alpha_\sigma} [\epsilon^\mu ({\bf p},\sigma )
]^\ast$, with $\alpha_\sigma$ being arbitrary phase factors at this stage.
Thus, ${\cal C} = I_{4\times 4}$ and $S^C ={\cal K}$.} \, We observe
that in the massless limit all the defined polarization vectors of the
momentum space do not have good behaviour; the functions describing spin-1
particles tend to infinity.\footnote{
It is interesting to remind that the authors of Ref.~\cite{Itzyk}
(see page 136 therein) tried to inforce the Stueckelberg's Lagrangian in
order to overcome the difficulties related with the $m\rightarrow 0$ limit
(or the Proca theory $\rightarrow$ Quantum Electrodynamics).  The
Stueckelberg's Lagrangian is well known to contain an additional term
which may be put in correspondence to some scalar (longitudinal) field
(cf.~Ref.~\cite{Staru}).} Nevertheless,
after renormalizing the potentials, {\it e.~g.}, $\epsilon^\mu \rightarrow
u^\mu \equiv m \epsilon^\mu$ we come to the field functions in the momentum
representation:
\begin{eqnarray}
&&u^\mu
({\bf p}, +1)= -{N\over \sqrt{2}m}\begin{pmatrix}p_r\cr m+ {p_1 p_r \over
E_p+m}\cr im +{p_2 p_r \over E_p+m}\cr {p_3 p_r \over
E_p+m}\end{pmatrix}\,,\,\\  &&u^\mu ({\bf p}, -1)= {N\over
\sqrt{2}m}\begin{pmatrix}p_l\cr m+ {p_1 p_l \over E_p+m}\cr -im +{p_2 p_l \over
E_p+m}\cr {p_3 p_l \over E_p+m}\end{pmatrix}\,,\label{vp12}
\end{eqnarray}
\begin{equation}
\quad \quad \quad u^\mu ({\bf
p}, 0) = {N\over m}\begin{pmatrix}p_3\cr {p_1 p_3 \over E_p+m}\cr {p_2 p_3
\over E_p+m}\cr m + {p_3^2 \over E_p+m}\end{pmatrix}\,,  \label{vp3}
\end{equation}
($N=m$ and $p_{r,l} = p_1 \pm ip_2$) which do not
diverge in the massless limit.  Two of the massless functions (with
 $\sigma = \pm 1$) are equal to zero when a particle, described by this
field, is moving along the third axis ($p_1 = p_2 =0$,\, $p_3 \neq 0$).
The third one ($\sigma = 0$) is
\begin{equation} u^\mu (p_3, 0)
\mid_{m\rightarrow 0} = \begin{pmatrix}p_3\cr 0\cr 0\cr {p_3^2 \over E_p}\end{pmatrix}
\equiv  \begin{pmatrix}E_p \cr 0 \cr 0 \cr E_p\end{pmatrix}\,, \end{equation}
\setcounter{footnote}{0}
and at
the rest ($E_p=p_3 \rightarrow 0$) also vanishes. Thus, such a field
operator describes the ``longitudinal photons" what is in the complete
accordance with the Weinberg theorem $B-A= h$
for massless particles (we
use the $D(1/2,1/2)$ representation). Thus, the change of the
normalization can lead to the ``change" of physical content described by
the classical field.
In the quantum case one should somehow fix the form of commutation
relations by some physical principles. They may be fixed by
requirements of the dimensionless of the action in the natural unit system 
(apart from the
requirements of the translational and rotational invariancies;
the accustomed behaviour of the Feynman-Dyson propagator), {\it etc.}

Furthermore, it is easy to find  the properties of the physical
fields $F^{\mu\nu}$ (defined as in (\ref{1},\ref{2}),
for instance) in the
massless zero-momentum limit. It is straightforward to find ${\bf
B}^{(+)} ({\bf p}, \sigma ) = {i\over 2m} {\bf p} \times {\bf u}
({\bf p}, \sigma)$, \, ${\bf E}^{(+)} ({\bf p}, \sigma) = {i\over 2m}
p_0 {\bf u} ({\bf p}, \sigma) - {i\over 2m} {\bf p} u^0
({\bf p}, \sigma)$ and the corresponding negative-energy strengths
for the field operator (in general, complex-valued)
\begin{eqnarray}
&& F^{\mu\nu}=\sum_{\sigma=0,\pm 1}
\int \frac{d^3 {\bf p}}{(2\pi)^3 2E_p} \, \left [
F^{\mu\nu}_{(+)} ({\bf p},\sigma)\, a ({\bf p},\sigma)\, e^{-ip x}\right .\nonumber\\
&&+ \left . F^{\mu\nu}_{(-)} ({\bf p},\sigma) \,b^\dagger ({\bf p},\sigma)\,
e^{+ip x} \right ]\,, \label{fop}
\end{eqnarray}
see Refs.~\cite{DVO961,DVO97}.

For the sake of completeness let us
present the vector corresponding to the ``time-like" polarization:
\begin{eqnarray}
&&u^\mu ({\bf p}, 0_t) = {N \over m} \begin{pmatrix}E_p\cr p_1
\cr p_2\cr p_3\end{pmatrix}\,,\,\\ 
&&{\bf B}^{(\pm)} ({\bf p}, 0_t) = {\bf
0}\,,\, {\bf E}^{(\pm)} ({\bf p}, 0_t) = {\bf 0}\,\,.
\label{tp}
\end{eqnarray}
The polarization vector $u^\mu ({\bf p}, 0_t)$ has
good behaviour in $m\rightarrow 0$, $N=m$ (and also in the subsequent
limit ${\bf p} \rightarrow {\bf 0}$) and it may correspond to some
field (particle).
As one can see the field operator  may describe a situation when a particle 
and an antiparticle
have {\it opposite} intrinsic parities, if it was composed of the state
of longitudinal polazrization (e.g., as the ``positive-energy" solution) 
and that of time-like polarization
(e.g., as the ``negative-energy" solution).  Furthermore, in the
case of the normalization of potentials to the mass $N=m$  the physical
fields ${\bf B}$ and ${\bf E}$, which correspond to the ``time-like"
polarization, are equal to zero identically.  The longitudinal fields
(strengths) are equal to zero in this limit only when one chooses the
frame  with $p_3 = \mid {\bf p} \mid$, cf. with the light front
formulation, Ref.~\cite{DVALF}.  In the case $N=1$ and (\ref{1},\ref{2})
the fields ${\bf B}^\pm ({\bf p},
0_t)$ and ${\bf E}^\pm ({\bf p}, 0_t)$ would be undefined.

\section{Lagrangian, Energy-Momentum Tensor and Angular Momentum.}

We begin with the Lagrangian,
including, in general, mass term:\footnote{Here we use the notation $A_{\mu\nu}$
for the AST due to different ``mass dimension" of the fields. The massless
($m=0$) Lagrangian is connected with the Lagrangians
used in the conformal field theories  by
adding the total derivative:
\begin{equation} {\cal L}_{CFT} = {\cal L} +
{1\over 2}\partial_\mu \left ( A_{\nu\alpha} \partial^\nu A^{\mu\alpha} -
A^{\mu\alpha} \partial^\nu A_{\nu\alpha} \right )\,.  \end{equation}
The Kalb-Ramond gauge-invariant
form (with
respect to ``gauge" transformations $A_{\mu\nu}  \rightarrow A_{\mu\nu}
+\partial_\nu \Lambda_\mu - \partial_\mu \Lambda_\nu$),
Refs.~\cite{Ogievet,Hayashi,Kalb}, is obtained only if one uses the Fermi
procedure {\it mutatis mutandis} by removing the additional ``phase" field
$\lambda (\partial_\mu A^{\mu\nu})^2$ from the Lagrangian. This has certain analogy with the QED,
where the question, whether the Lagrangian is gauge-invariant or not, is
solved depending on the presence of the term $\lambda (\partial_\mu
A^\mu)^2$. For details see Refs.~\cite{Hayashi,Kalb}.

In general it is possible to introduce various forms of the mass term
and forms of corresponding normalization of the field. But, the dimensionless
of the action ${\cal S}$  imposes some restrictions. We know that
$A^{\mu\nu}$ (in order to be able to describe long-range forces) should have
the dimension $[energy]^2$ in the natural unit system. 
In order to take this into account one should
divide the Lagrangian (\ref{Lagran}) by $m^2$; calculate
corresponding dynamical invariants, other observable quantities; and only
then study $m\rightarrow 0$ limit.}
\begin{eqnarray} {\cal L} &=&  {1\over
4} (\partial_\mu A_{\nu\alpha})(\partial^\mu A^{\nu\alpha}) - {1\over 2}
(\partial_\mu A^{\mu\alpha})(\partial^\nu A_{\nu\alpha}) - \nonumber\\
&-& {1\over 2}
(\partial_\mu A_{\nu\alpha})(\partial^\nu A^{\mu\alpha}) + 
{1\over 4} m^2
A_{\mu\nu} A^{\mu\nu} \,.  \label{Lagran} \end{eqnarray} The Lagrangian
leads to the equation of motion in the following form (provided that the
appropriate antisymmetrization procedure has been taken into account):
\begin{equation} {1\over 2} ({\,\lower0.9pt\vbox{\hrule
\hbox{\vrule height 0.2 cm \hskip 0.2 cm \vrule height
0.2cm}\hrule}\,}+m^2) A_{\mu\nu} +
(\partial_{\mu}A_{\alpha\nu}^{\,,\alpha} -
\partial_{\nu}A_{\alpha\mu}^{\,,\alpha}) = 0 \,,\label{PE}
\end{equation}
where ${\,\lower0.9pt\vbox{\hrule \hbox{\vrule height 0.2 cm
\hskip 0.2 cm
\vrule height 0.2 cm}\hrule}\,}
=- \partial_{\alpha}\partial^{\alpha}$, cf. with the set of equations
(15,16).  It is this equation for antisymmetric-tensor-field components
that follows from the Proca-Duffin-Kemmer
consideration
provided that $m\neq 0$ and in the final expression one takes into account
the Klein-Gordon equation $({\,\lower0.9pt\vbox{\hrule \hbox{\vrule height
0.2 cm \hskip 0.2 cm \vrule height 0.2 cm}\hrule}\,} - m^2) A_{\mu\nu}=
0$.  The latter expresses relativistic dispersion relations $E^2 -{\bf p}^2
=m^2$.

Following the variation procedure
one can obtain the energy-momentum tensor:
\begin{eqnarray}
\Theta^{\lambda\beta} &=& {1\over 2} \left [
(\partial^\lambda A_{\mu\alpha}) (\partial^\beta A^{\mu\alpha})
- 2(\partial_\mu A^{\mu\alpha}) (\partial^\beta A^\lambda_{\,\alpha}) -
\right.\nonumber\\
&-& \left . 2 (\partial^\mu A^{\lambda\alpha}) (\partial^\beta
A_{\mu\alpha})\right ] -{\cal L} g^{\lambda\beta}\, .
\end{eqnarray}
One can also obtain that
for rotations $x^{\mu^\prime} = x^\mu + \omega^{\mu\nu} x_\nu$
the corresponding variation of the wave function is found
from the formula:
\begin{equation}
\delta A^{\alpha\beta} = {1\over 2} \omega^{\kappa\tau}
{\cal T}_{\kappa\tau}^{\alpha\beta,\mu\nu} A_{\mu\nu}\,.
\end{equation}
The generators of infinitesimal transformations are then defined as
\begin{eqnarray}
\lefteqn{{\cal T}_{\kappa\tau}^{\alpha\beta,\mu\nu} \,=\,
{1\over 2} g^{\alpha\mu} (\delta_\kappa^\beta \,\delta_\tau^\nu \,-\,
\delta_\tau^\beta\,\delta_\kappa^\nu) \,+\,{1\over 2} g^{\beta\mu}
(\delta_\kappa^\nu\delta_\tau^\alpha  \,-\,
\delta_\tau^\nu\, \delta_\kappa^\alpha) \nonumber}\\
&&+
{1\over 2} g^{\alpha\nu} (\delta_\kappa^\mu \, \delta_\tau^\beta \,-\,
\delta_\tau^\mu \,\delta_\kappa^\beta) \,+\, {1\over 2}
g^{\beta\nu} (\delta_\kappa^\alpha \,\delta_\tau^\mu \,-\,
\delta_\tau^\alpha \, \delta_\kappa^\mu).
\end{eqnarray}
It is ${\cal T}_{\kappa\tau}^{\alpha\beta,\mu\nu}$, the generators of
infinitesimal transformations,
that enter in the formula for the relativistic spin tensor:
\begin{equation}
J_{\kappa\tau} = \int d^3 {\bf x} \left [ \frac{\partial {\cal
L}}{\partial ( \partial A^{\alpha\beta}/\partial t )} {\cal
T}^{\alpha\beta,\mu\nu}_{\kappa\tau} A_{\mu\nu} \right ]\,.
\label{inv}
\end{equation}
As a result one  obtains:
\begin{eqnarray}
J_{\kappa\tau} &=& \int d^3 {\bf x} \left [ (\partial_\mu A^{\mu\nu})
(g_{0\kappa} A_{\nu\tau} - g_{0\tau} A_{\nu\kappa}) - \right .\nonumber\\
&-&\left . (\partial_\mu
A^\mu_{\,\,\,\,\kappa}) A_{0\tau} + (\partial_\mu A^\mu_{\,\,\,\,\tau})
A_{0\kappa} + \right. \nonumber\\
&+& \left. A^\mu_{\,\,\,\,\kappa} ( \partial_0 A_{\tau\mu} +
\partial_\mu A_{0\tau} +\partial_\tau A_{\mu 0})  \right . \nonumber\\
&-&  \left . A^\mu_{\,\,\,\,\tau}
( \partial_0 A_{\kappa\mu} +\partial_\mu A_{0\kappa} +\partial_\kappa
A_{\mu 0}) \right ]\,. \label{gene10}
\end{eqnarray}
If one agrees that the
orbital part of the angular momentum
\begin{equation} L_{\kappa\tau} =
x_\kappa \Theta_{0\,\tau} - x_\tau \Theta_{0\,\kappa} \,,
\end{equation}
with  $\Theta_{\tau\lambda}$ being the energy-momentum tensor, does not
contribute to the Pauli-Lubanski operator when acting on the
one-particle free states (as in the Dirac $s=1/2$ case), then
the Pauli-Lubanski 4-vector is constructed as
follows, Eq. (2-21) of Ref.~\cite{Itzyk} :
\begin{equation}
W_\mu = -{1\over 2}  \epsilon_{\mu\kappa\tau\nu} J^{\kappa\tau} P^\nu \,,
\end{equation}
with $J^{\kappa\tau}$ defined by Eqs.
(\ref{inv},\ref{gene10}). The 4-momentum operator $P^\nu$ can be replaced
by its eigenvalue when acting on the plane-wave eigenstates.

Furthermore, one should
choose space-like normalized vector $n^\mu n_\mu = -1$, for example $n_0
=0$,\, ${\bf n} = \widehat  {\bf p} = {\bf p} /\vert {\bf
p}\vert$.\,\,\footnote{One should remember that the helicity operator is
usually connected with the Pauli-Lubanski vector in the following manner
$({\bf J} \cdot \widehat {\bf p}) = ({\bf W} \cdot \widehat {\bf p})/
E_p$, see Ref.~\cite{Shirok}. The choice of Ref.~\cite{Itzyk}, p. 147,
$n^\mu = \left ( t^\mu - p^\mu {p\cdot t \over m^2} \right ) {m\over \mid
{\bf p} \mid}$, with $t^\mu \equiv (1,0,0,0)$ being a time-like vector, is
also possible but it leads to some oscurities in the procedure of taking
the massless limit.} \,\, After lengthy calculations in a spirit of pp.
58, 147 of Ref.~\cite{Itzyk}, one can find the explicit form of
the relativistic spin:
\begin{eqnarray} && (W_\mu \cdot n^\mu) = - ({\bf
W}\cdot {\bf n}) = -{1\over 2} \epsilon^{ijk} n^k J^{ij}
p^0\,,\label{PL1}\\
&& {\bf J}^k = {1\over 2} \epsilon^{ijk} J^{ij} =
\epsilon^{ijk} \int d^3 {\bf x} \left [ A^{0i} (\partial_\mu A^{\mu j}) + \right .\nonumber\\
&+& \left . A_\mu^{\,\,\,\,j} (\partial^0 A^{\mu i} +\partial^\mu A^{i0} +\partial^i
A^{0\mu} ) \right ]\,.\label{PL2}
\end{eqnarray}
Now it becomes obvious that the application of the generalized Lorentz
conditions (which are the quantum versions of free-space dual Maxwell's
equations) leads in such a formulation to the absence of electromagnetism
in a conventional sense.  The resulting Kalb-Ramond field is longitudinal
(helicity $h=0$).  All the components of the angular momentum tensor
for this case are identically equated to zero.

According to~\cite[Eqs.(9,10)]{Ogievet} we proceed in the
construction of the ``potentials" for the notoph :
\begin{equation}
\tilde F_{\mu\nu} ({\bf p}) \sim A_{\mu\nu} ({\bf p}) = N \left [\epsilon_\mu^{(1)} ({\bf
p})\epsilon_\nu^{(2)} ({\bf p})- \epsilon_\nu^{(1)} ({\bf p})
\epsilon_\mu^{(2)} ({\bf p}) \right ] \end{equation}
On using explicit
forms for the polarization vectors in the momentum space  one obtains
\begin{eqnarray}
&&A^{\mu\nu} = {iN^2 \over m} \begin{pmatrix}0&-p_2&
p_1& 0\cr p_2 &0& m+{p_r p_l\over p_0+m} & {p_2 p_3\over p_0 +m}\cr -p_1 &
-m - {p_r p_l \over p_0+m}& 0& -{p_1 p_3\over p_0 +m}\cr 0& -{p_2 p_3
\over p_0 +m} & {p_1 p_3 \over p_0+m}&0\end{pmatrix} \nonumber\\
&&\label{lc} \end{eqnarray}
i.e., it coincides with the longitudinal components of the antisymmetric
tensor obtained in Refs.~\cite[Eqs.(2.14,2.17)]{Ahlu}
and~\cite[Eqs.(17b,18b)]{DVO97}  within the normalization and
different forms of the spin basis.  The
longitudinal states reduce to zero in the massless case under appropriate
choice of the normalization and only if a $s=1$ particle moves along with
the third axis $OZ$. 

Finally, we agree with the previous authors, e.~g., Ref.~\cite{Ohanian} ,
see Eq. (4) therein, about the gauge {\it non}-invariance of
the separation of the angular momentum of the electromagnetic field into the
``orbital" and ``spin" part (\ref{PL2}). We proved again
that for the antisymmetric tensor field ${\bf J} \sim \int d^3 {\bf x}\,
({\bf E}\times {\bf A})$. So, what people actually did (when spoken about
the Ogievetski\u{\i}-Polubarinov-Kalb-Ramond field) is:
When $N=m$ they considered the gauge part of the 4-vector field functions.
Then, they equated ${\bf A}$ of the transverse modes on choosing
$p_r =p_l =0$ in the massless limit
(see formulas (\ref{vp12})).\footnote{The reader, of course,
can consider this procedure as the usual gauge transformation, $A^\mu
\rightarrow A^\mu +\partial^\mu \chi$.} Under this choice the ${\bf E}
({\bf p}, 0)$ and ${\bf B} ({\bf p}, 0)$ are equal to zero in massless
limit.  But, the gauge part of $u^\mu ({\bf p}, 0)$ is not. The spin
angular momentum can still be zero.

\section{The Relations with the $2(2s+1)$ Formalism. Photon-Notoph Equations.}

For the spin 1 one can start from
\begin{equation}
[\gamma_{\alpha\beta} p^\alpha p^\beta - A p^\alpha p_\alpha +Bm^2] \Psi
=0\,, \end{equation} where $p_\mu=-i\partial_\mu$ and
$\gamma_{\alpha\beta}$ are the Barut-Muzinich-Williams covariantly-defined 
$6\times 6$ matrices.
One can consider four cases:
\begin{itemize}

\item
$\Psi^{(I)} = \begin{pmatrix}{\bf E} +i{\bf B}\cr
{\bf E} -i{\bf B}\end{pmatrix}$, $P=-1$, where ${\bf E}_i$ and ${\bf B}_i$ are the
components of the tensor.

\item
$\Psi^{(II)} = \begin{pmatrix}{\bf B} -i{\bf E}\cr
{\bf B} +i{\bf E}\end{pmatrix}$, $P=+1$, where ${\bf
E}_i$, ${\bf B}_i$ are the components of the tensor.

\item
$\Psi^{(III)} = \Psi^{(I)}$, but ${\bf E}_i$ and ${\bf B}_i$
are the
corresponding vector and axial-vector  components of the
{\it dual} tensor $\tilde F_{\mu\nu}$.

\item
$\Psi^{(IV)} = \Psi^{(II)}$, where ${\bf E}_i$ and ${\bf B}_i$
are the components of the {\it dual} tensor $\tilde F_{\mu\nu}$.

\end{itemize}
The mappings of the Weinberg-Tucker-Hammer (WTH) equations are:
\begin{eqnarray}
&&\partial_\alpha\partial^\mu F_{\mu\beta}^{(I)}
-\partial_\beta\partial^\mu F_{\mu\alpha}^{(I)}\nonumber\\
&+& {A-1\over 2} \partial^\mu \partial_\mu F_{\alpha\beta}^{(I)}
+{B\over 2} m^2 F_{\alpha\beta}^{(I)} = 0\,,\label{wth1}\\
&&\partial_\alpha\partial^\mu F_{\mu\beta}^{(II)}
-\partial_\beta\partial^\mu F_{\mu\alpha}^{(II)}\nonumber\\
&-& {A+1\over 2} \partial^\mu \partial_\mu F_{\alpha\beta}^{(II)}
-{B\over 2} m^2 F_{\alpha\beta}^{(II)} = 0\,,\\
&&\partial_\alpha\partial^\mu \tilde F_{\mu\beta}^{(III)}
-\partial_\beta\partial^\mu \tilde F_{\mu\alpha}^{(III)}\nonumber\\
&-& {A+1\over 2} \partial^\mu \partial_\mu \tilde F_{\alpha\beta}^{(III)}
-{B\over 2} m^2 \tilde F_{\alpha\beta}^{(III)} = 0\,,\\
&&\partial_\alpha\partial^\mu \tilde F_{\mu\beta}^{(IV)}
-\partial_\beta\partial^\mu \tilde F_{\mu\alpha}^{(IV)}\nonumber\\
&+& {A-1\over 2} \partial^\mu \partial_\mu \tilde F_{\alpha\beta}^{(IV)}
+{B\over 2} m^2 \tilde F_{\alpha\beta}^{(IV)} = 0\,,
\end{eqnarray}
where the superindices $(I)-(IV)$ refer to different forms of the WTH field functions , which are composed from ${\bf E}$
and ${\bf B}$, polar and axial 3-vectors.
In the Tucker-Hammer case ($A=1, B=2$) we can recover the Proca theory
from (\ref{wth1}):
\begin{equation}
\partial_\alpha \partial^\mu F_{\mu\beta}
-\partial_\beta \partial^\mu F_{\mu\alpha} + m^2 F_{\alpha\beta} =0
\label{proca1}\,,
\end{equation}
($A_\nu =-{1\over m^2}  \partial^\alpha F_{\alpha\nu}$ should be substituted in
$F_{\mu\nu} = \partial_\mu A_\nu -\partial_\nu A_\mu$, and multiplied by
$m^2$).

We also noted that the massless limit of this theory contains the Maxwell theory as a particular
case.  In~\cite{DVO961,DVO97,physscripta} we showed that it is possible to define
various massless limits for the Proca-Duffin-Kemmer theory. Another one is
the Ogievetski\u{\i}-Polubarinov {\it notoph}, see above. The transverse
components of the AST field can be removed from the corresponding
Lagrangian by means of the ``new gauge transformation" $A_{\mu\nu}
\rightarrow A_{\mu\nu} +\partial_\mu \Lambda_\nu -\partial_\nu
\Lambda_\mu$, with the vector gauge function $\Lambda_\mu$.

Bargmann and Wigner claimed
explicitly that they constructed $(2s+1)$ states (the
Weinberg-Tucker-Hammer theory has  essentially $2(2s+1)$
components). Therefore, we now apply
\begin{eqnarray}
&&\Psi_{\{\alpha\beta\}} = (\gamma^\mu R)_{\alpha\beta} (c_a m A_\mu +
c_f  F_\mu) +\nonumber\\
&+& (\sigma^{\mu\nu} R)_{\alpha\rho} (c_A m (\gamma^5)_{\rho\beta}
A_{\mu\nu} + c_F I_{\rho\beta} F_{\mu\nu})\, .\label{si}
\end{eqnarray}
Thus, $A_\mu$, $A_{\mu\nu}$ and $F_\mu$, $F_{\mu\nu}$ have different mass dimension.
The constants $c_i$ are some numerical dimensionless coefficients. 
The $\gamma^{\mu} R$, $\sigma^{\mu\nu} R$
and $\gamma^5 \sigma^{\mu\nu} R$ are the {\it symmetrical} matrices.
The substitution of the above expansion into the Bargmann-Wigner
set, Ref.~\cite{Lurie},
gives us the new Proca-like equations:
\begin{eqnarray}
&&c_a m (\partial_\mu A_\nu - \partial_\nu A_\mu ) +
c_f (\partial_\mu F_\nu -\partial_\nu F_\mu ) =\nonumber\\
&&=ic_A m^2 \epsilon_{\alpha\beta\mu\nu} A^{\alpha\beta} +
2 m c_F F_{\mu\nu}\label{pr1}\\
&&c_a m^2 A_\mu + c_f m F_\mu =
i c_A m \epsilon_{\mu\nu\alpha\beta} \partial^\nu A^{\alpha\beta} +
2 c_F \partial^\nu F_{\mu\nu}\,\nonumber\\
&& . \label{pr2}
\end{eqnarray}
In the case $c_a=1$, $c_F ={1\over 2}$ and $c_f=c_A=0$ they
are reduced to the ordinary Proca equations.
In the general case we obtain
dynamical equations which connect the photon, the notoph and
their potentials. The divergent (in $m\rightarrow 0$) parts
of field functions and those of dynamical variables
should be removed by the corresponding gauge (or the Kalb-Ramond gauge)
transformations. It is  known that the notoph massless field is
considered to be the pure
longitudinal field ($h=0$) after one takes into account $\partial_\mu
A^{\mu\nu}=
0$.  Apart from these dynamical equations we can obtain a number of
constraints by means of subtraction of the equations of the
Bargmann-Wigner system (instead of addition as for
(\ref{pr1},\ref{pr2})). In fact, they give 
$\widetilde F^{\mu\nu} \sim im A^{\mu\nu}$
and $F^\mu \sim mA^\mu$, as in Ref.~\cite{Ogievet} .
Thus, after the suitable choice of the dimensionless coefficients
$c_i$ the
Lagrangian density for the photon-notoph field can be proposed:
\begin{eqnarray}
{\cal L} &=& {\cal L}^{Proca} +{\cal L}^{Notoph} = -
{1\over 8} F_\mu F^\mu -{1\over 4} F_{\mu\nu} F^{\mu\nu} +\nonumber\\
&+& {m^2 \over 2} A_\mu A^\mu + {m^2 \over 4} A_{\mu\nu} A^{\mu\nu}\, ,
\end{eqnarray}
The limit $m\rightarrow 0$ may be taken for dynamical variables,
in the end of calculations only.

Furthermore, it is logical to introduce the normalization scalar field
$\varphi (x)$ and to consider the expansion:
\begin{equation}
\Psi_{\{\alpha\beta\}} = (\gamma^\mu R)_{\alpha\beta} (\varphi A_\mu)
+ (\sigma^{\mu\nu} R)_{\alpha\beta} F_{\mu\nu}\, .
\end{equation}
Then, we arrive at the following set
\begin{eqnarray}
&&2m F_{\mu\nu} = \varphi (\partial_\mu A_\nu -\partial_\nu A_\mu)
+ (\partial_\mu \varphi) A_\nu - (\partial_\nu \varphi) A_\mu\, ,\nonumber\\
&&\\
&& \partial^\nu F_{\mu\nu} = {m \over 2} (\varphi A_\mu)\, ,
\end{eqnarray}
which in the case of the constant
scalar field $\varphi = 2m$  can again be reduced to the system of the
Proca equations. The additional constraints are
\begin{eqnarray}
&&(\partial^\mu \varphi) A_\mu + \varphi (\partial^\mu A_\mu) =0\,,\\
&& \partial_\mu \widetilde F^{\mu\nu} = 0\, .
\end{eqnarray}
 
At this moment, it is not yet obvious, how can we
account for other equations in the $(1,0)\oplus (0,1)$
representation rigorously.   One can wish to seek the generalization of
the Proca system on the basis of the introduction of two mass parameters
$m_1$ and $m_2$.   Another equation in the
$(1/2,0)\oplus (0,1/2)$ representation was discussed in
Ref.~\cite{Raspini} :
\begin{equation} \left [
i\gamma^\mu \partial_\mu - m_1 - \gamma^5 m_2 \right ] \Psi (x) =0\,.
\end{equation}
The Bargmann-Wigner procedure for this system of 
the equations (which include the $\gamma^5$ matrix in the mass term) gives:
\begin{eqnarray}
&&2m_1  F^{\mu\nu} +2i m_2 \widetilde F^{\mu\nu} = \varphi
(\partial^\mu A^\nu -\partial^\nu A^\mu) +\nonumber\\
&+&(\partial^\mu \varphi) A^\nu - (\partial^\nu \varphi) A^\mu\, ,\\
&&\partial^\nu F_{\mu\nu} = {m_1 \over 2} (\varphi A_\mu), \,
\end{eqnarray}
with the constraints
\begin{eqnarray}
&&(\partial^\mu \varphi) A_\mu +\varphi (\partial^\mu A_\mu) = 0\,, \\
&&\partial^\nu \widetilde F_{\mu\nu} = {im_2 \over 2} (\varphi A_\mu)\,.
\end{eqnarray}

Next, the Tam-Happer experiments~\cite{TH} on two laser beams interaction did not
find satisfactory explanation in the framework of the ordinary QED.
On the other hand, in Ref.~\cite{Pradhan} a very interesting model has
been proposed.  It is based on
gauging the Dirac field on using the coordinate-dependent parameters
$\alpha_{\mu\nu} (x)$ in
\begin{equation}
\psi(x) \rightarrow \psi^\prime (x^\prime) = \Omega \psi(x)\,\,, \,
\Omega = \exp \left [ {i\over 2} \sigma^{\mu\nu} \alpha_{\mu\nu}(x)
\right ]\, .
\end{equation}
Thus,  the second ``photon" was introduced. The
compensating 24-component field $B_{\mu,\nu\lambda}$
reduces
to the 4-vector field as follows:
\begin{equation}
B_{\mu,\nu\lambda} = {1\over 4} \epsilon_{\mu\nu\lambda\sigma} a^\sigma
(x) \, .
\end{equation}
As readily seen after comparison of these formulas with those of
Refs.~\cite{Ogievet,Hayashi,Kalb} , the second photon is nothing more than the
Ogievetski\u{\i}-Polubarinov {\it notoph} within the normalization.

\section{The Bargmann-Wigner Formalism  for Spin 2.}

In this Section we use the commonly-accepted procedure
for the derivation  of higher-spin equations~\cite{BW}. 
We begin with the equations for the 4-rank symmetric spinor:
\begin{eqnarray}
\left [ i\gamma^\mu \partial_\mu - m \right ]_{\alpha\alpha^\prime}
\Psi_{\alpha^\prime \beta\gamma\delta} &=& 0\, ,\\
\left [ i\gamma^\mu \partial_\mu - m \right ]_{\beta\beta^\prime}
\Psi_{\alpha\beta^\prime \gamma\delta} &=& 0\, ,\\
\left [ i\gamma^\mu \partial_\mu - m \right ]_{\gamma\gamma^\prime}
\Psi_{\alpha\beta\gamma^\prime \delta} &=& 0\, ,\\
\left [ i\gamma^\mu \partial_\mu - m \right ]_{\delta\delta^\prime}
\Psi_{\alpha\beta\gamma\delta^\prime} &=& 0\, .
\end{eqnarray}  
We proceed expanding the field function in the complete set of symmetric matrices. 
In the beginning let us use the
first two indices:
\begin{equation} \Psi_{\{\alpha\beta\}\gamma\delta} =
(\gamma_\mu R)_{\alpha\beta} \Psi^\mu_{\gamma\delta}
+(\sigma_{\mu\nu} R)_{\alpha\beta} \Psi^{\mu\nu}_{\gamma\delta}\, .
\label{bwff}
\end{equation}
We would like to write
the corresponding equations for functions $\Psi^\mu_{\gamma\delta}$
and $\Psi^{\mu\nu}_{\gamma\delta}$ in the form:
\begin{eqnarray}
&&{2\over m} \partial_\mu \Psi^{\mu\nu}_{\gamma\delta} = -
\Psi^\nu_{\gamma\delta}\, , \label{p1}\\
&&\Psi^{\mu\nu}_{\gamma\delta} = {1\over 2m}
\left [ \partial^\mu \Psi^\nu_{\gamma\delta} - \partial^\nu
\Psi^\mu_{\gamma\delta} \right ]\, \label{p2}.
\end{eqnarray}  
The constraints $(1/m) \partial_\mu \Psi^\mu_{\gamma\delta} =0$
and \linebreak $(1/m) \epsilon^{\mu\nu}_{\,\alpha\beta}\, \partial_\mu
\Psi^{\alpha\beta}_{\gamma\delta} = 0$ can be regarded as a consequence
of
Eqs.  (\ref{p1},\ref{p2}). 
Next, we present the vector-spinor and tensor-spinor functions as
\begin{eqnarray}
&&\Psi^\mu_{\{\gamma\delta\}} = (\gamma^\kappa R)_{\gamma\delta}
G_{\kappa}^{\, \mu} +(\sigma^{\kappa\tau} R )_{\gamma\delta}
F_{\kappa\tau}^{\, \mu} \, ,\\
&&\Psi^{\mu\nu}_{\{\gamma\delta\}} = (\gamma^\kappa R)_{\gamma\delta}
T_{\kappa}^{\, \mu\nu} +(\sigma^{\kappa\tau} R )_{\gamma\delta}
R_{\kappa\tau}^{\, \mu\nu} \, ,
\end{eqnarray} 
i.~e.,  using the symmetric matrices  in indices $\gamma$ and
$\delta$. Hence,  the resulting tensor equations are (cf.~Ref.~\cite{MS}) :
\begin{eqnarray}
&&{2\over m} \partial_\mu T_\kappa^{\, \mu\nu} =
-G_{\kappa}^{\,\nu}\, ,\\
&&{2\over m} \partial_\mu R_{\kappa\tau}^{\, \mu\nu} =
-F_{\kappa\tau}^{\,\nu}\, ,\\
&& T_{\kappa}^{\, \mu\nu} = {1\over 2m} \left [
\partial^\mu G_{\kappa}^{\,\nu}
- \partial^\nu G_{\kappa}^{\, \mu} \right ] \, ,\\
&& R_{\kappa\tau}^{\, \mu\nu} = {1\over 2m} \left [
\partial^\mu F_{\kappa\tau}^{\,\nu}
- \partial^\nu F_{\kappa\tau}^{\, \mu} \right ] \, .
\end{eqnarray}  
The constraints are re-written to
\begin{eqnarray}
&&{1\over m} \partial_\mu G_\kappa^{\,\mu} = 0\, ,\,
{1\over m} \partial_\mu F_{\kappa\tau}^{\,\mu} =0\, ,\\
&& {1\over m} \epsilon_{\alpha\beta\nu\mu} \partial^\alpha
T_\kappa^{\,\beta\nu} = 0\, ,\,
{1\over m} \epsilon_{\alpha\beta\nu\mu} \partial^\alpha
R_{\kappa\tau}^{\,\beta\nu} = 0\, .\nonumber\\
&&
\end{eqnarray}
 
However, we need to make symmetrization over these two sets
of indices $\{ \alpha\beta \}$ and $\{\gamma\delta \}$. The total
symmetry can be ensured if one contracts the function
$\Psi_{\{\alpha\beta
\} \{\gamma \delta \}}$ with the {\it antisymmetric} matrices
$R^{-1}_{\beta\gamma}$, $(R^{-1} \gamma^5 )_{\beta\gamma}$ and
$(R^{-1} \gamma^5 \gamma^\lambda )_{\beta\gamma}$, and equate
all these contractions to zero (similar to the $s=3/2$ case
considered in Ref.~\cite[p. 44]{Lurie}). We obtain
additional constraints on the tensor field functions.
We explicitly showed that all field functions become to be equal to
zero.
Such a situation cannot be considered as a satisfactory one, because it
does not give us any physical information.

We shall modify the formalism in the spirit of  Ref.~\cite{physscripta}.
The field function (\ref{bwff}) is now presented as
\begin{eqnarray}
&&\Psi_{\{\alpha\beta\}\gamma\delta} =
\alpha_1 (\gamma_\mu R)_{\alpha\beta} \Psi^\mu_{\gamma\delta} +
\alpha_2 (\sigma_{\mu\nu} R)_{\alpha\beta} \Psi^{\mu\nu}_{\gamma\delta}+\nonumber\\
&&+\alpha_3 (\gamma^5 \sigma_{\mu\nu} R)_{\alpha\beta}
\widetilde \Psi^{\mu\nu}_{\gamma\delta}\, ,
\end{eqnarray}
with
\begin{eqnarray}
&&\Psi^\mu_{\{\gamma\delta\}} = \beta_1 (\gamma^\kappa R)_{\gamma\delta}
G_\kappa^{\,\mu} + \beta_2 (\sigma^{\kappa\tau} R)_{\gamma\delta}
F_{\kappa\tau}^{\,\mu} +\nonumber\\
&+&\beta_3 (\gamma^5 \sigma^{\kappa\tau}
R)_{\gamma\delta} \widetilde F_{\kappa\tau}^{\,\mu} \, ,\\
&&\Psi^{\mu\nu}_{\{\gamma\delta\}} =\beta_4 (\gamma^\kappa
R)_{\gamma\delta} T_\kappa^{\,\mu\nu} + \beta_5 (\sigma^{\kappa\tau}
R)_{\gamma\delta} R_{\kappa\tau}^{\,\mu\nu} +\nonumber\\
&+&\beta_6 (\gamma^5
\sigma^{\kappa\tau} R)_{\gamma\delta}
\widetilde R_{\kappa\tau}^{\,\mu\nu} \, ,\\
&&\widetilde \Psi^{\mu\nu}_{\{\gamma\delta\}} =\beta_7 (\gamma^\kappa
R)_{\gamma\delta} \widetilde T_\kappa^{\,\mu\nu} + \beta_8
(\sigma^{\kappa\tau} R)_{\gamma\delta}
\widetilde D_{\kappa\tau}^{\,\mu\nu}+\nonumber\\
&+&\beta_9 (\gamma^5 \sigma^{\kappa\tau} R)_{\gamma\delta}
D_{\kappa\tau}^{\,\mu\nu} \, .
\end{eqnarray}
 
Hence, the function $\Psi_{\{\alpha\beta\}\{\gamma\delta\}}$
can be expressed as a sum of nine terms.
The corresponding dynamical
equations are given in the following form:
  \begin{eqnarray}
&& {2\alpha_2
\beta_4 \over m} \partial_\nu T_\kappa^{\,\mu\nu} +{i\alpha_3
\beta_7 \over m} \epsilon^{\mu\nu\alpha\beta} \partial_\nu
\widetilde T_{\kappa,\alpha\beta} = \alpha_1 \beta_1
G_\kappa^{\,\mu} \label{b}\\
&&{2\alpha_2 \beta_5 \over m} \partial_\nu
R_{\kappa\tau}^{\,\mu\nu} +{i\alpha_2 \beta_6 \over m}
\epsilon_{\alpha\beta\kappa\tau} \partial_\nu \widetilde R^{\alpha\beta,
\mu\nu} +\nonumber\\
&&+{i\alpha_3 \beta_8 \over m}
\epsilon^{\mu\nu\alpha\beta}\partial_\nu \widetilde
D_{\kappa\tau,\alpha\beta} - 
{\alpha_3 \beta_9 \over 2}
\epsilon^{\mu\nu\alpha\beta} \epsilon_{\lambda\delta\kappa\tau}
D^{\lambda\delta}_{\, \alpha\beta} = \nonumber\\
&&=\alpha_1 \beta_2
F_{\kappa\tau}^{\,\mu} + {i\alpha_1 \beta_3 \over 2}
\epsilon_{\alpha\beta\kappa\tau} \widetilde F^{\alpha\beta,\mu}\,, \\
&& 2\alpha_2 \beta_4 T_\kappa^{\,\mu\nu} +i\alpha_3 \beta_7
\epsilon^{\alpha\beta\mu\nu} \widetilde T_{\kappa,\alpha\beta}=\nonumber\\
&&=  {\alpha_1 \beta_1 \over m} (\partial^\mu G_\kappa^{\, \nu}
- \partial^\nu G_\kappa^{\,\mu})\,, \\
&& 2\alpha_2 \beta_5 R_{\kappa\tau}^{\,\mu\nu} +i\alpha_3 \beta_8
\epsilon^{\alpha\beta\mu\nu} \widetilde D_{\kappa\tau,\alpha\beta}+\nonumber\\
&&+i\alpha_2 \beta_6 \epsilon_{\alpha\beta\kappa\tau} \widetilde
R^{\alpha\beta,\mu\nu}
- {\alpha_3 \beta_9\over 2} \epsilon^{\alpha\beta\mu\nu}
\epsilon_{\lambda\delta\kappa\tau} D^{\lambda\delta}_{\, \alpha\beta}
= \nonumber\\
&&= {\alpha_1 \beta_2 \over m} (\partial^\mu F_{\kappa\tau}^{\, \nu}
-\partial^\nu F_{\kappa\tau}^{\,\mu} ) + \nonumber\\
&&+{i\alpha_1 \beta_3 \over 2m}
\epsilon_{\alpha\beta\kappa\tau} (\partial^\mu \widetilde
F^{\alpha\beta,\nu} - \partial^\nu \widetilde F^{\alpha\beta,\mu} )\, .
\label{f}
\end{eqnarray} 
In general, the coefficients $\alpha_i$ and $\beta_i$
may now carry some dimension.
The essential constraints can be found in Ref.~\cite{DvoeglazovAACA}.
They are  the results of contractions of the field function
with six antisymmetric matrices, as above. 
As a discussion, we note that in such a framework we have
physical
content because only certain combinations of field functions
can be equal to zero. In general, the fields
$F_{\kappa\tau}^{\,\mu}$, $\widetilde F_{\kappa\tau}^{\,\mu}$,
$T_{\kappa}^{\,\mu\nu}$, $\widetilde T_{\kappa}^{\,\mu\nu}$, and
$R_{\kappa\tau}^{\,\mu\nu}$,  $\widetilde
R_{\kappa\tau}^{\,\mu\nu}$, $D_{\kappa\tau}^{\,\mu\nu}$,
$\widetilde
D_{\kappa\tau}^{\,\mu\nu}$ correspond to different physical
states.
The equations  describe couplings one state to another.

Furthermore, from the set of equations (\ref{b}-\ref{f}) one
obtains the {\it second}-order equation for symmetric traceless tensor
of
the second rank ($\alpha_1 \neq 0$, $\beta_1 \neq 0$):
\begin{equation} {1\over m^2} \left [\partial_\nu
\partial^\mu G_\kappa^{\, \nu} - \partial_\nu \partial^\nu
G_\kappa^{\,\mu} \right ] =  G_\kappa^{\, \mu}\, .\label{geq}
\end{equation}
After the contraction in indices $\kappa$ and $\mu$ this equation is
reduced to:
\begin{eqnarray}
&&\partial_\mu G_{\,\kappa}^{\mu} = F_\kappa\,  \\
&&{1\over m^2} \partial_\kappa F^\kappa = 0\, ,
\end{eqnarray}
i.~e., the equations which connect the analogue of the energy-momentum
tensor and the analogue of the 4-vector potential. 
See also the works on the notivarg concept~\cite{Tybor}.  
Further investigations may provide additional foundations to
``surprising" similarities of gravitational and electromagnetic
equations in the low-velocity limit, Refs.~\cite{Wein2,Logunov,KG,Rodrigues}.

\section{Interactions with Fermions.}

The possibility of terms such as  $\mathbf\sigma\cdot [ {\mathbf A} \times {\mathbf A}^\ast ]$ appears to be related to the matters
of chiral interactions~\cite{DVO-UNUSUAL,DVO-APS}. As we are now convinced, the Dirac field operator 
can be always
presented as a superposition of the self- and anti-self charge conjugate field operators (cf.~Ref.~\cite{Ziino}). The
anti-self charge conjugate part gives the self charge conjugate part after multiplying by
the $\gamma^5$ matrix, and {\it vice versa}. We derived
\begin{equation}
[ i\gamma^\mu D^\ast_\mu - m ]\psi^s_1 = 0 \, ,\label{31}
\end{equation}
\begin{equation}
[ i\gamma^\mu D_\mu - m ]\psi^a_2 = 0\,.\label{32}
\end{equation}
Both equations lead to the  interaction terms such as ${\mathbf\sigma}\cdot [ {\mathbf A} \times {\mathbf A}^\ast ]$ provided that the
4-vector potential is considered as a complex function(al). In fact, from (\ref{31}) we have:
\begin{equation}
i\sigma^\mu \nabla_\mu \chi_1 -  m\phi_1 = 0\,,\,
i\tilde\sigma^\mu \nabla_\mu^\ast \phi_1 -  m\chi_1 = 0\,.
\end{equation}
And, from (\ref{32}) we have
\begin{equation}
i\sigma^\mu \nabla_\mu^\ast \chi_2 -  m\phi_2 = 0\,,\,
i\tilde\sigma^\mu \nabla_\mu \phi_2 -  m\chi_2 = 0\,.
\end{equation}
The meanings of $\sigma^\mu$ and $\tilde\sigma^\mu$ are obvious from the definition of $\gamma$ matrices. 
The derivatives are defined:
\begin{equation}
D_\mu =\partial_\mu -ie\gamma^5 C_\mu +eB_\mu\,,\,
\nabla_\mu =\partial_\mu -ieA_\mu\,,
\end{equation}
and $A_\mu =C_\mu +iB_\mu$. Thus, relations with the magnetic monopoles can be established.

From the above systems we extract the terms $\pm e^2 \sigma^i \sigma^j A_i A^\ast_j$, 
which lead to the discussed terms~\cite{DVO-UNUSUAL,DVO-APS}.
We would like to note that  the terms of the type ${\boldsymbol\sigma}\cdot [ {\mathbf A} \times {\mathbf A}^\ast ]$ can be reduced
to $({\boldsymbol\sigma}\cdot {\mathbf\nabla}) V$, where $V$ is the scalar potential.

Furthermore, one can come to the same conclusions not applying to the constraints on the
creation/annihilation operators (which we  previously choose for clarity and simplicity in~\cite{DVO-APS}). 
It is possible to work with self/anti-self charge conjugate fields and  the Majorana
{\it anzatzen}.
Thus, it is the $\gamma^5$ transformation which distinguishes various
field configurations (helicity, self/anti-self charge conjugate properties etc) in the coordinate
representation in the considered cases.

\section{Boson Interactions.}

The most general relativistic-invariant Lagrangian for the symmetric 2nd-rank tensor is
\begin{eqnarray}
\lefteqn{ {\cal L} = -\alpha_1 (\partial^\alpha G_{\alpha\lambda}) (\partial_\beta G^{\beta\lambda})
-\alpha_2 (\partial_\alpha G^{\beta\lambda}) (\partial^\alpha G_{\beta\lambda})}-\nonumber\\
&-&\alpha_3 (\partial^\alpha G^{\beta\lambda}) (\partial_\beta G_{\alpha\lambda})
+m^2 G_{\alpha\beta} G^{\alpha\beta}\,.\label{lagran}
\end{eqnarray} 
It leads to the equation
\begin{equation}
\left [ \alpha_2 (\partial_\alpha \partial^\alpha) +m^2 \right ] G^{\left \{\mu\nu\right\}} + (\alpha_1 +\alpha_3) \partial^{\left \{\mu \vert\right.}
(\partial_\alpha G^{\left.\alpha \vert \nu \right\}})=0\,.
\end{equation}
In the case $\alpha_2 =1 \textgreater 0$ and $\alpha_1+\alpha_3=-1$ it coincides with Eq. (\ref{geq}).
There is no any problem to obtain the dynamical invariants for the fields of the spin 2 from the above Lagrangian.
The mass dimension of $G^{\mu\nu}$ is $[energy]^1$.
We now present possible relativistic interactions of the symemtric 2nd-rank tensor. 
The simplest ones should be the the following ones:
\begin{eqnarray}
{\cal L}^{int}_{(1)} &\sim& G_{\mu\nu} F^\mu F^\nu\,,\\
{\cal L}^{int}_{(2)}  &\sim& (\partial^\mu G_{\mu\nu}) F^\nu\,,\\
{\cal L}^{int}_{(3)} &\sim& G_{\mu\nu} (\partial^\mu F^\nu)\,.
\end{eqnarray}
The term $(\partial_\mu G^\alpha_{\,\,\,\alpha}) F^\mu$ vanishes due to the constraint
of tracelessness.  The interaction with the notoph can be related o the scalar-tensor  
theories of gravity.

It is also interesting to note that thanks to the possible terms 
\begin{equation}
V (F) =\lambda_1 (F_\mu F^\mu) + \lambda_2 (F_\mu F^\mu)(F_\nu F^\nu)
\end{equation}
we can give the mass to the $G_{00}$ component of the spin-2 field. This is due to
the possibility of the Higgs spontaneous symmetry breaking~\cite{Higgs}:
\begin{equation}
F^\mu (x) =\begin{pmatrix}v+\partial_0 \chi (x)\cr g^1\cr g^2\cr g^3\end{pmatrix}\,,
\end{equation}
with $v$ being the vacuum expectation value, $v^2 =(F_\mu F^\mu)= -\lambda_1/2\lambda_2 \textgreater 0$.
Other degrees of freedom of the 4-vector field are removed since they are the Goldstone 
bosons. It was stated that ``for any continuous
symmetry which does not preserve the ground state, there is a massless degree of freedom 
which decouples at low energies. This mode is called the Goldstone (or Nambu-Goldstone) particle
for the symmetry". As usual, the Goldstone modes  should be important in
giving masses to the three vector bosons.
As one can easily seen, this expression does not permit an arbitrary phase for $F^\mu$, which is only possible if 
the 4-vector would be the complex one.

Next, since the interaction of fermions with notoph, for instance, are that of the order $\sim e^2$ since the beginning
(as opposed to the fermion current interaction with the 4-vector potential $A_\mu$) in the Lagrangian, 
it is more difficult to observe it.
However, as far as I know, the theoretical precision calculus in QED (the Land\'e factor, the anomalous magnetic moment, the hyperfine splittings in positronium and muonium, and the decay rate of {\it o-Ps} and {\it p-Ps})  are near  the order corresponding to the 4th-5th loops, where the difference may appear with the experiments, cf.~\cite{DFT,Kinoshita}.

\section{Conclusions}

We considered the Bargmann-Wigner formalism in order to derive the equations for the AST fields, and for 
the symmetric tensor of the 2nd rank.
We introduced the additional normalization scalar field in the Bargmann-Wigner formalism in order 
to account for possible physical
significance of the Ogievetskii-Polubarinov--Kalb-Ramond modes. Both the  antisymmetric tensor fields and
the 4-vector fields may have third helicity state in the massless limit
from a theoretical viewpoint.
This problem is connected with the problem of the observability of
the gauge~\cite{Staru}.  
We introduced the additional symmetric matrix in the
Bargmann-Wigner expansion $(\gamma^5 \sigma^{\mu\nu} R)$ in order to account for the dual fields.
The consideration was similar to Ref.~\cite{DVO-FS}. 
The problem was discussed, what are the  
the correct definitions of the energy-momentum tensor and other N\"other currents
in the electromagnetic theory,
the relativistic theory of gravitation, the general relativity, and their generalizations.
Furthermore, we discussed the interactions of notoph, photon and graviton. Probably, 
the notivarg should also be taken into account. In order to analize its dynamical invariants and interactions
one should construct Lagrangian from the analogs of the Riemann tensor, such as $\widetilde D^{\mu\nu,\alpha\beta}$.
For instance, the interaction notoph-graviton may give the mass to spin-2 particles in the way similar to 
the spontaneous-symmetry-breaking Higgs formalism.

\section*{Acknowledgments}

I am grateful to the referee of ``International Journal of Modern Physics" and
``Foundation of Physics", whose advice of the mass dimension
(normalization) of the fields  was very useful.
I acknowledge discussions with participants of recent conferences
on Symmetries.  

\balance

\end{document}